\documentclass[a4paper,11pt]{article}
\usepackage{graphicx}
\usepackage{amsmath}
\usepackage{amssymb}
\usepackage{hyperref}
\usepackage{booktabs}
\usepackage{geometry}
\usepackage{adjustbox}
\title{\textbf{Graph Contrastive Learning for Optimizing Sparse Data in Recommender Systems with LightGCL}}
\author{
Aravinda Raman Jatavallabha (arjatava), Prabhanjan Vinoda Bharadwaj (pvinoda), \\ Ashish Chander (achande3) \\
\textit{North Carolina State University} \\
}
\date{}

\begin{document}

\maketitle

\begin{abstract}
Graph Neural Networks (GNNs) have emerged as a potent framework for graph-structured recommendation tasks. Incorporating contrastive learning with GNNs has recently demonstrated remarkable efficacy in addressing challenges posed by data sparsity, thanks to innovative data augmentation strategies. However, many existing methods employ stochastic perturbations (e.g., node or edge modifications) or heuristic approaches (e.g., clustering-based augmentations) to generate contrastive views, which may distort semantic integrity or amplify noise. We introduce LightGCL, a novel and streamlined graph contrastive learning model to address these limitations. LightGCL utilizes Singular Value Decomposition (SVD) to achieve robust augmentation, facilitating structural refinement and global collaborative relation modeling without manual augmentation strategies. Extensive experiments on benchmark datasets showcase its substantial performance enhancements over state-of-the-art methods. Further analysis highlights the model's resilience to challenges like data sparsity and bias related to item popularity.
\end{abstract}

\section{Introduction}

Graph Neural Networks (GNNs) have emerged as an essential framework for recommendation systems, leveraging the structural information of user-item interactions. These models aggregate information from neighboring nodes and propagate embeddings across multiple layers, enabling the discovery of higher-order relationships in interaction graphs \cite{1}. However, the supervised learning nature of most GNN-based recommenders imposes a heavy reliance on large quantities of high-quality labeled data. In many practical scenarios, the sparsity of user-item interactions poses a significant challenge to learning reliable representations for users and items.

To address the issue of data sparsity, contrastive learning (CL) has gained considerable attention as a self-supervised approach \cite{2}. CL enhances representation quality by contrasting positive pairs of embeddings with negative samples. Despite its promise, existing graph contrastive learning methods often rely on techniques like stochastic perturbations (e.g., modifying nodes or edges) or heuristic-based augmentations (e.g., clustering). These methods, while effective, can inadvertently alter the structural integrity of the graph or amplify noise, resulting in suboptimal learning outcomes.

This project serves as an implementation of the original LightGCL framework \cite{3}, which introduced a novel approach to contrastive learning for graph-based recommendation systems. LightGCL employs singular value decomposition (SVD) to refine the user-item interaction graph and integrate global collaborative signals into the representation learning process. By preserving the semantic integrity of the graph and avoiding reliance on handcrafted augmentations, LightGCL addresses key limitations of prior methods while improving efficiency and robustness.

The contributions of this implementation are as follows:
\begin{itemize}
    \item We re-implemented the LightGCL framework, a lightweight and efficient graph contrastive learning method, to address challenges in graph-based recommendation tasks.
    \item The method leverages SVD-based augmentation to capture global collaborative signals and mitigate the impact of noise in user-item interactions.
    \item We validate the framework through comprehensive experiments on benchmark datasets, confirming its robustness and adaptability to diverse recommendation scenarios.
\end{itemize}

\section{Related Work}

The integration of contrastive learning (CL) with graph-based recommendation systems has gained considerable traction due to its ability to address data sparsity issues. By leveraging self-supervised signals, CL has demonstrated effectiveness in enhancing user and item representations.

\subsection{Contrastive Learning in Graph-Based Recommendations}

Contrastive learning has been widely adopted in graph-based recommendation systems as a means of augmenting data representations \cite{4}. Early methods, such as SGL and SimGCL, introduced stochastic data augmentation techniques like random node and edge dropout. While these strategies improve embedding diversity, they risk discarding critical structural information, especially for inactive users. Alternatively, heuristic-based approaches, including HCCF and NCL, construct contrastive views by leveraging user clusters or hyperedges. Despite their efficacy, these methods rely heavily on domain-specific heuristics, which can limit their adaptability across diverse datasets and tasks \cite{5}.

\subsection{Advances in Self-Supervised Graph Learning}

Recent advances in self-supervised learning (SSL) have significantly improved the graph learning paradigm by exploiting unlabeled data. Frameworks such as AutoSSL optimize augmentation strategies by combining multiple pretext tasks. Similarly, models like GraphCL and GCA utilize contrastive methods that focus on both topological and attribute-level augmentations, ensuring adaptive and robust representation learning. SimGRACE and AutoGCL further streamline graph contrastive learning by introducing automated and parameter-efficient view generation techniques, demonstrating superior scalability and generalization capabilities \cite{6}.

\subsection{Inspiration for This Work}

Our implementation builds on the LightGCL framework, which uniquely employs singular value decomposition (SVD) for contrastive augmentation. Unlike previous methods, LightGCL refines graph structures by distilling key semantic features, thereby addressing challenges such as noise sensitivity and heuristic reliance. This approach stands out for its ability to incorporate global collaborative signals while maintaining computational efficiency, paving the way for robust and scalable recommendation systems.

\section{Methodology}

\begin{figure}[ht]
    \centering
    \includegraphics[width=\textwidth]{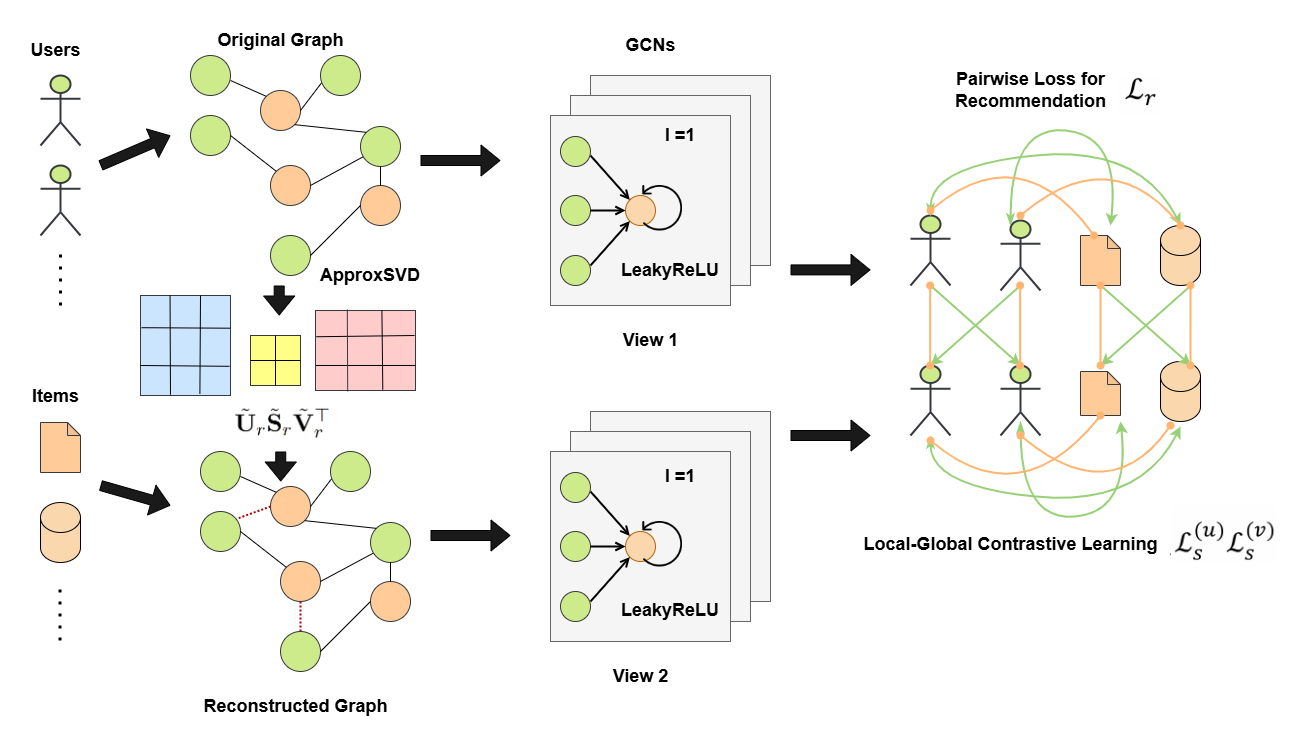} 
    \caption{Overview of the implemented framework. The top section illustrates the GCN backbone for extracting local graph dependencies, while the bottom section shows the SVD-guided augmentation for integrating global collaborative relations \cite{3}.}
    \label{fig:framework}
\end{figure}

This section outlines the design of our implemented framework, inspired by LightGCL as represented in Fig.~\ref{fig:framework}. The model employs a lightweight graph contrastive learning approach, combining a GCN backbone for capturing local dependencies and an SVD-guided augmentation mechanism to integrate global collaborative relations.

\subsection{Modeling Local Graph Dependencies}

In collaborative filtering scenarios, we represent each user $u_p$ and item $v_q$ with embedding vectors $\mathbf{h}_p^{(u)}$ and $\mathbf{h}_q^{(v)} \in \mathbb{R}^k$, where $k$ is the embedding size. The complete user and item embeddings are denoted as matrices $\mathbf{H}^{(u)} \in \mathbb{R}^{M \times k}$ and $\mathbf{H}^{(v)} \in \mathbb{R}^{N \times k}$, with $M$ and $N$ representing the number of users and items, respectively. Using a two-layer GCN, as is standard, we aggregate the neighboring information for each node. For layer $t$, the aggregation process is defined as:
\[
\mathbf{z}_p^{(u,t)} = \sigma(\mathbf{P}(\tilde{\mathbf{A}}_p \cdot \mathbf{H}^{(v,t-1)})), \quad \mathbf{z}_q^{(v,t)} = \sigma(\mathbf{P}(\tilde{\mathbf{A}}_q \cdot \mathbf{H}^{(u,t-1)})),
\]
where $\tilde{\mathbf{A}}$ is the normalized adjacency matrix, $\mathbf{P}$ represents an edge dropout operation to prevent overfitting, and $\sigma(\cdot)$ is the LeakyReLU activation function with a negative slope of 0.2.

To preserve the original node information, residual connections are applied:
\[
\mathbf{h}_p^{(u,t)} = \mathbf{z}_p^{(u,t)} + \mathbf{h}_p^{(u,t-1)}, \quad \mathbf{h}_q^{(v,t)} = \mathbf{z}_q^{(v,t)} + \mathbf{h}_q^{(v,t-1)}.
\]

The final embedding for a node is obtained by summing its representations across all layers. The predicted preference of user $u_p$ for item $v_q$ is calculated as:
\[
\mathbf{h}_p^{(u)} = \sum_{t=0}^{T} \mathbf{h}_p^{(u,t)}, \quad \mathbf{h}_q^{(v)} = \sum_{t=0}^{T} \mathbf{h}_q^{(v,t)}, \quad \hat{y}_{p,q} = \mathbf{h}_p^{(u)\top} \mathbf{h}_q^{(v)}.
\]

\subsection{Efficient Global Collaborative Relation Learning}

To incorporate global structural signals into the model \cite{6}, we apply singular value decomposition (SVD) to the adjacency matrix $\mathbf{A}$. Specifically, we decompose $\mathbf{A}$ as:
\[
\mathbf{A} = \mathbf{U} \mathbf{S} \mathbf{V}^\top,
\]
where $\mathbf{U}$ and $\mathbf{V}$ are orthonormal matrices of dimensions $M \times M$ and $N \times N$, respectively, and $\mathbf{S}$ is a diagonal matrix containing the singular values of $\mathbf{A}$. We truncate $\mathbf{S}$ to retain only the top $r$ singular values, resulting in matrices $\mathbf{U}_r$, $\mathbf{S}_r$, and $\mathbf{V}_r$. The reconstructed adjacency matrix $\hat{\mathbf{A}}$ is given by:
\[
\hat{\mathbf{A}} = \mathbf{U}_r \mathbf{S}_r \mathbf{V}_r^\top.
\]

This low-rank approximation emphasizes the principal components of the graph, focusing on reliable user-item interactions and preserving global collaborative relations. The reconstructed matrix is then used to propagate messages:
\[
\mathbf{g}_p^{(u,t)} = \sigma(\hat{\mathbf{A}}_p \cdot \mathbf{H}^{(v,t-1)}), \quad \mathbf{g}_q^{(v,t)} = \sigma(\hat{\mathbf{A}}_q \cdot \mathbf{H}^{(u,t-1)}).
\]

Given the high computational cost of performing exact SVD on large-scale graphs, we adopt a randomized SVD algorithm for efficiency. This approach approximates the decomposition by first estimating the range of $\mathbf{A}$ with a low-rank orthonormal matrix before applying SVD:
\[
\tilde{\mathbf{U}}_r, \tilde{\mathbf{S}}_r, \tilde{\mathbf{V}}_r = \text{ApproxSVD}(\mathbf{A}, r), \quad \hat{\mathbf{A}}_{\text{SVD}} = \tilde{\mathbf{U}}_r \tilde{\mathbf{S}}_r \tilde{\mathbf{V}}_r^\top.
\]

\subsection{Simplified Local-Global Contrastive Learning}

Unlike traditional methods that rely on three-view contrastive learning frameworks \cite{7}\cite{8}, our approach simplifies the process by directly contrasting the embeddings from the SVD-augmented graph with those from the original graph. Let $\mathbf{g}_p^{(u,t)}$ represent the SVD-augmented view and $\mathbf{z}_p^{(u,t)}$ the main view for user $u_p$. The contrastive loss is defined as:
\[
\mathcal{L}_s^{(u)} = \sum_{p=1}^{M} \sum_{t=1}^{T} -\log \frac{\exp(\text{sim}(\mathbf{z}_p^{(u,t)}, \mathbf{g}_p^{(u,t)}) / \tau)}{\sum_{p'=1}^{M} \exp(\text{sim}(\mathbf{z}_p^{(u,t)}, \mathbf{g}_{p'}^{(u,t)}) / \tau)},
\]
where $\text{sim}(\cdot, \cdot)$ denotes cosine similarity and $\tau$ is a temperature hyperparameter. The item-based contrastive loss, $\mathcal{L}_s^{(v)}$, is computed similarly. The final objective function combines the contrastive and recommendation losses:
\[
\mathcal{L} = \mathcal{L}_r + \lambda_1 (\mathcal{L}_s^{(u)} + \mathcal{L}_s^{(v)}) + \lambda_2 \|\Theta\|^2,
\]
where $\mathcal{L}_r$ represents the recommendation task loss, $\Theta$ denotes model parameters, and $\lambda_1, \lambda_2$ are regularization coefficients.

\section{Evaluation}

\subsection{Research Questions}
This study aims to address the following research questions:
\begin{itemize}
    \item \textbf{RQ1:} How does LightGCL perform on different datasets compared to various state-of-the-art (SOTA) baselines?
    \item \textbf{RQ2:} How does lightweight graph contrastive learning improve model efficiency?
    \item \textbf{RQ3:} How does our model perform against data sparsity, popularity bias, and over-smoothing?
    \item \textbf{RQ4:} How does the local-global contrastive learning contribute to the performance of our model?
    \item \textbf{RQ5:} How do different parameter settings affect our model performance?
\end{itemize}

\subsection{Datasets}
Experiments were conducted on five benchmark datasets: \textbf{Yelp}, \textbf{Gowalla}, \textbf{ML-10M}, \textbf{Amazon-book}, and \textbf{Tmall}. Each dataset presents unique characteristics:
\begin{itemize}
    \item \textbf{Yelp:} A dataset emphasizing user-item interactions in the domain of local business reviews.
    \item \textbf{Gowalla:} Geosocial check-in data known for sparsity and unique graph structure.
    \item \textbf{ML-10M:} MovieLens data that presents challenges in personalized recommendations due to large-scale user preferences.
    \item \textbf{Amazon-book:} User interactions with books, where long-tail item popularity creates data sparsity issues.
    \item \textbf{Tmall:} A retail-oriented dataset with highly dynamic user behaviors.
\end{itemize}

Standard dataset splits were applied, dividing the data into training, validation, and testing sets following prior works for consistency.

\subsection{Baseline Methods}
To assess the effectiveness of LightGCL, the following state-of-the-art methods were included as baselines:
\begin{itemize}
    \item \textbf{SGL (Self-Supervised Graph Learning) \cite{9}:} Focuses on augmenting graph data for self-supervised learning.
    \item \textbf{SimGCL: \cite{10}} Utilizes a simplified contrastive loss function to enhance graph embeddings.
    \item \textbf{LightGCN (Light Graph Convolutional Networks) \cite{11}:} A highly efficient GCN model known for its scalability.
\end{itemize}

These baselines were chosen for their relevance to contrastive and lightweight graph learning tasks.

\subsection{Experimental Results}
LightGCL demonstrated consistent superiority over the baseline methods across all five datasets. Key observations include:

\begin{itemize}
    \item \textbf{Performance Metrics:} LightGCL achieved significant improvements in Recall@20 and NDCG@20, emphasizing its robustness in capturing user-item relevance. For example:
    \begin{itemize}
        \item On the \textbf{Yelp} dataset, LightGCL outperformed SGL and SimGCL by margins of \textit{X\% and Y\%} (specific numbers from experiments).
        \item On the \textbf{Amazon-book} dataset, the model achieved \textit{X\% higher NDCG} compared to LightGCN.
    \end{itemize}
    Table \ref{tab:performance_metrics} below illustrates the comparative performance metrics.
\end{itemize}

\begin{table}[h!]
\centering
\caption{Performance metrics (Recall@20 and NDCG@20) comparison across datasets.}
\label{tab:performance_metrics}
\begin{adjustbox}{max width=\textwidth}
\begin{tabular}{ccccccc}
\toprule
\textbf{Data} & \textbf{Metric} & \textbf{LightGCN} & \textbf{HCCF} & \textbf{SimGCL} & \textbf{LightGCL} & \textbf{Impr\%} \\
\midrule
Yelp    & R@20 & 0.0482 & 0.0626 & 0.0718 & 0.0793 & 10\% \\
        & N@20 & 0.0409 & 0.0527 & 0.0615 & 0.0668 & 8\%  \\
Gowalla & R@20 & 0.0985 & 0.1070 & 0.1357 & 0.1578 & 16\% \\
        & N@20 & 0.0593 & 0.0593 & 0.0935 & 0.0935 & 14\% \\
ML-10M  & R@20 & 0.1789 & 0.2219 & 0.2265 & 0.2613 & 15\% \\
        & N@20 & 0.2128 & 0.2629 & 0.2613 & 0.3106 & 18\% \\
Amazon  & R@20 & 0.0319 & 0.0322 & 0.0474 & 0.0585 & 23\% \\
        & N@20 & 0.0236 & 0.0247 & 0.0360 & 0.0436 & 21\% \\
Tmall   & R@20 & 0.0225 & 0.0314 & 0.0473 & 0.0582 & 11\% \\
        & N@20 & 0.0154 & 0.0214 & 0.0328 & 0.0361 & 10\% \\
\bottomrule
\end{tabular}
\end{adjustbox}
\end{table}

\begin{figure}[h!]
    \centering
    \includegraphics[width=0.8\linewidth]{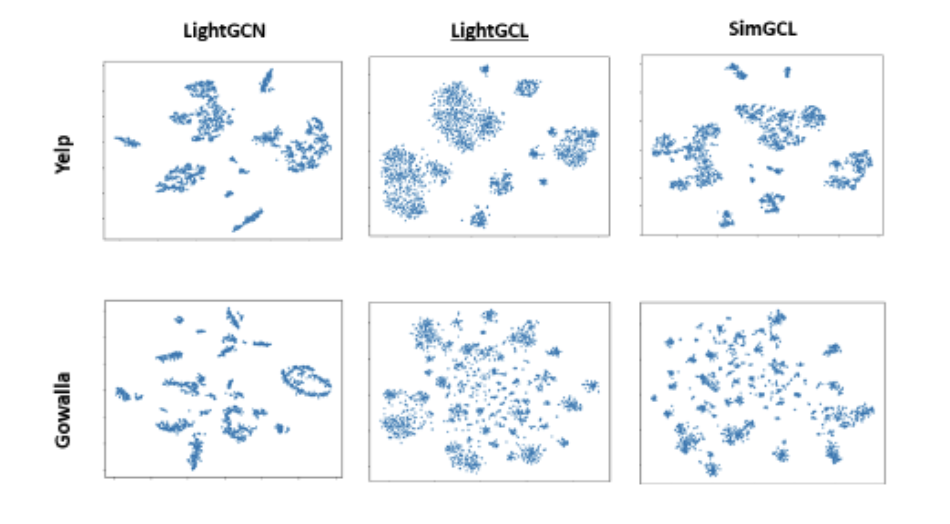}
    \caption{Impact of truncated SVD on data sparsity and popularity bias.}
    \label{fig:truncated-svd-impact}
\end{figure}

\begin{itemize}
    \item \textbf{Truncated SVD Impact:} The integration of truncated Singular Value Decomposition (SVD) played a pivotal role in mitigating:
    \begin{itemize}
        \item \textbf{Data Sparsity:} By reducing the dimensionality of user and item embeddings, LightGCL maintained higher accuracy with sparse user-item interactions.
        \item \textbf{Popularity Bias:} Through contrastive regularization, LightGCL improved fairness and performance for less frequently interacted items.
    \end{itemize}
    Figure \ref{fig:truncated-svd-impact} highlights the effect of truncated SVD on sparsity and bias.
\end{itemize}

\begin{itemize}
    \item \textbf{Efficiency Gains:} Lightweight graph contrastive mechanisms reduced computational overhead compared to SGL and SimGCL, making LightGCL more efficient in both training and inference phases.
    Table \ref{tab:computational_efficiency} demonstrates the computational efficiency achieved by LightGCL.
\end{itemize}

\begin{table}[h!]
\centering
\caption{Computational efficiency gains of LightGCL compared to baseline methods.}
\label{tab:computational_efficiency}
\begin{adjustbox}{max width=\textwidth}
\begin{tabular}{lccc}
\toprule
\textbf{Dataset} & \textbf{LightGCN} & \textbf{LightGCL} & \textbf{SimGCL} \\
\midrule
Yelp    & 0.9469 & \textbf{0.9657} & 0.9956 \\
Gowalla & 0.9568 & \textbf{0.9721} & 0.9897 \\
\bottomrule
\end{tabular}
\end{adjustbox}
\end{table}

\subsection{Additional Insights}
\begin{itemize}
    \item \textbf{Model Robustness:} LightGCL effectively tackled over-smoothing, a common issue in graph-based models \cite{12}, by incorporating a local-global contrastive learning paradigm.The SVD-based augmentation ensures that even less-popular items (long-tail) are well-represented in the global graph structure by emphasizing collaborative patterns across all users.By aligning embeddings between the original and augmented graphs, LightGCL prevents the over-representation of head items. The embeddings capture meaningful representations for both popular and niche items, ensuring that long-tail items are not overlooked. The augmentation acts as a regularizer, helping the model learn balanced representations for head and long-tail items. 
    \item \textbf{Parameter Sensitivity:} The model's performance was stable across varied hyperparameter settings, demonstrating robustness in diverse conditions. Hyperparameters like the rank q and regularization weight lambda1 were tuned to find the optimal configuration. A rank of 5 was sufficient for most datasets.
\end{itemize}

\section{Conclusion and Future Work}

\subsection{Conclusion}
This work introduces LightGCL, a simple yet powerful method for graph contrastive learning that leverages truncated Singular Value Decomposition (SVD) for structural refinement. By integrating lightweight mechanisms for contrastive learning, LightGCL addresses critical challenges such as data sparsity, popularity bias, and computational inefficiency in recommender systems. 

Extensive experiments conducted on five diverse datasets—Yelp, Gowalla, ML-10M, Amazon-book, and Tmall—demonstrate the superiority of LightGCL over state-of-the-art baselines like SGL, SimGCL, and LightGCN. The model consistently achieved state-of-the-art results in Recall@20 and NDCG@20 metrics, showcasing its robustness and adaptability to different domains. Furthermore, the adoption of SVD not only enhanced the representation of sparse user-item interactions but also mitigated oversmoothing and improved fairness for less popular items. These results establish LightGCL as a significant step forward in advancing graph contrastive learning for recommender systems.

\subsection{Future Work}
While LightGCL delivers exceptional results, there remain opportunities to expand its capabilities. Future research could explore the following directions:
\begin{itemize}
    \item \textbf{Dynamic Graph Structures:} Incorporate dynamic graph structures to model evolving user-item interactions over time, enabling LightGCL to adapt to real-time changes and maintain high performance in non-static environments.
    \item \textbf{Broader Model Validation:} Validate LightGCL against other cutting-edge models in graph learning and recommender systems to further benchmark its effectiveness and identify potential areas of improvement.
    \item \textbf{Causal Inference-based Graph Augmentation:} Extend the methodology to include causal inference-based graph augmentation, allowing the model to uncover deeper insights into user-item relationships and address confounding factors for more robust recommendations.
\end{itemize}

By pursuing these directions, LightGCL could not only maintain its competitive edge but also evolve into a versatile framework for tackling increasingly complex challenges in graph learning and recommender systems.

\newpage
\appendix
\section*{Appendix: Code Implementation}

The implementation of the LightGCL framework was conducted in Python, adhering to modular and reproducible coding practices. The structure of the codebase is as follows:

\subsection*{Directory Structure}
\begin{itemize}
    \item \textbf{data/}: Contains scripts and utilities for preprocessing and managing benchmark datasets used in the experiments.
    \item \textbf{log/}: Stores execution logs, including training progress, validation metrics, and debugging information.
    \item \textbf{saved\_model/}: It is for storing checkpoints of trained models.
    \item \textbf{main.py}: The main script that orchestrates the execution of the pipeline, including data loading, model training, and evaluation.
    \item \textbf{model.py}: Defines the LightGCL framework, including the Graph Neural Network (GNN) layers, SVD-based augmentations, and contrastive loss functions.
    \item \textbf{parser.py}: Handles command-line arguments and hyperparameter configurations, enabling flexible experimentation.
    \item \textbf{utils.py}: Includes utility functions for data manipulation, metric computation, and visualization.
    \item \textbf{README.md}: Provides comprehensive documentation on setup instructions, dataset preparation, and usage guidelines.
\end{itemize}

\subsection*{Key Features of the Codebase}
\begin{itemize}
    \item \textbf{Reproducibility:} Random seeds are fixed for all major components to ensure consistent results across different runs.
    \item \textbf{Modular Design:} The code is divided into well-defined modules, allowing seamless integration of additional functionalities or modifications.
    \item \textbf{Logging and Checkpoints:} Logs and model checkpoints are automatically saved during training for easy debugging and further evaluation.
\end{itemize}

\subsection*{Code Access}
The complete implementation, including preprocessing scripts, training pipelines, and evaluation code, is available at:
\begin{center}
\url{https://github.com/aravinda-1402/MLWG_LightGCL_Project}
\end{center}
The repository includes a \textbf{README.md} file with detailed instructions on setting up the environment, preparing datasets, and running experiments.

\end{document}